\documentclass[reprint,
superscriptaddress,
 amsmath,amssymb,
 aps,
]{revtex4-1}

\usepackage{graphicx}
\usepackage{dcolumn}
\usepackage{bm}
\usepackage{physics}
\usepackage{hyperref}
\usepackage[usenames,dvipsnames]{color}
\usepackage{xspace}
\usepackage{siunitx}
\usepackage{multirow}

\usepackage{tikzsymbols}
\begin{document}

\newcommand{\hlc}[2][yellow]{ {\sethl$Q_o=9.7\times10^7$ and or{#1} \hl{#2}} }
\newcommand{\jcomment}[1]{\textcolor{ForestGreen}{Jev says: #1 } }
\newcommand{\rcomment}[1]{\textcolor{RoyalBlue}{Rose says: #1 } }
\definecolor{mpink}{RGB}{212,0,170}
\newcommand{\mcomment}[1]{\textcolor{mpink}{Matt says: #1 \Cat} }
\newcommand{\jono}[1]{\textcolor{Red}{Jono says: #1 } }
\newcommand{\citeit}[1]{[Cite: #1,\cite{#1}]}
\newcommand{\isoeuform}{Eu$^{35}$Cl\ensuremath{_3\cdot}6H\ensuremath{_2}O\xspace}
\newcommand{\euform}{EuCl\ensuremath{_3\cdot}6H\ensuremath{_2}O\xspace}
\newcommand{\erform}{ErCl\ensuremath{_3\cdot}6H\ensuremath{_2}O\xspace}
\newcommand{\yso}{Y\ensuremath{_2}SiO\ensuremath{_5}\xspace}

\preprint{APS/123-QED}

\title{Microwave to optical photon conversion via fully concentrated rare-earth ion crystals}

\author{Jonathan R. Everts}
\affiliation{%
 Dodd-Walls Centre for Photonic and Quantum Technologies, Department of Physics, University of Otago, Dunedin 9016, New Zealand.
}%

\author{Matthew C. Berrington}
\affiliation{
  Centre for Quantum Computation and Communication Technology, Research School of Physics and Engineering, The Australian National University, Canberra, Australian Capital Territory 0200, Australia
}

\author{Rose L. Ahlefeldt}
\affiliation{
  Centre for Quantum Computation and Communication Technology, Research School of Physics and Engineering, The Australian National University, Canberra, Australian Capital Territory 0200, Australia
}

\author{Jevon J. Longdell}%
\affiliation{%
 Dodd-Walls Centre for Photonic and Quantum Technologies, Department of Physics, University of Otago, Dunedin 9016, New Zealand.
}%

\date{\today}%

\begin{abstract}
  Most investigations of rare earth ions in solids for quantum information have used rare earth ion doped crystals. Here we analyse the conversion of quantum information from microwave photons to optical frequencies using crystals where the rare earth ions, rather than being dopants, are part of the host crystal. The potential of large ion densities and small linewidths makes such systems very attractive in this application. We show that, as well as high efficiency, large bandwidth conversion is possible. In fact, the collective coupling between the rare earth ions and the optical and microwave cavities is large enough that the limitation on the bandwidth of the devices will instead be the spacing between magnon mode modes in the crystal.
\end{abstract}

\maketitle

\section{\label{sec:intro}Introduction}
Over recent years, superconducting qubits have emerged as a leading qubit design for quantum information processing \cite{Nakamura1999, Devoret2013, You2005}. In these systems, quantum information can readily be coupled into and out of the qubits via microwave photons. However,  this means that they need to be operated at temperatures in the milli-kelvin range in order to not be swamped by thermal noise. It also makes long distance communication problematic because transferring quantum information with microwave photons would require a cryogenically cooled channel. Transferring quantum information to optical frequencies using a microwave-to-optical up-converter would get around this problem and enable the use of existing optical network technology.

To be useful for quantum computing, a microwave-to-optical up-converter needs nearly unity conversion efficiency for single photons and no added noise. This requires a system with a strong interaction with both optical and microwave fields, and a very strong nonlinearity to perform the conversion. Several different approaches have been investigated.   Electro-optomechanical approaches \cite{Barzanjeh2012, Wang2012, Bochmann2013, Bagci2014, Zhang2015, Lecocq2016} have achieved the highest  efficiency so far of 47\%, but the conversion bandwidth was limited to $12$\,kHz and there were 38 photons of added noise \cite{Andrews2014, Higginbotham2018}. Conversely, electro-optic materials \cite{Tsang10, Tsang11, Schewfel2016, Witmer2017, Soltani2017} have demonstrated a 2\% conversion efficiency but with a much larger bandwidth, of 0.59\,MHz at 2\,K  \cite{Fan2018}.

Approaches using atoms \cite{Gard2017,Verdu2009} have been proposed. Recently, using the strong microwave and optical transitions of  a cloud of Rydberg atoms at microkelvin temperatures, a conversion efficiency of 0.3\% has been achieved with a 4~MHz bandwidth \cite{Han2018}. Collective spin excitations within a ferromagnet (magnons) have also been suggested. These resonances interact strongly with microwave cavity fields because of the very high density of spins. Low efficiency up-conversion has been demonstrated in  a ferromagnetic crystal of yttrium iron garnet (YIG), limited by the weak coupling of the magnons to the optical field \cite{Nakamura2014, Zhang2014, Hu2016, Goryachev2014}.

Atom-like systems in solids have been investigated including defects in diamond \cite{Marcos2010,Kubo2010} and rare-earth doped solids \cite{Williamson2014, Obrien2014, Fernandez2015, Fernandez2017}.  In these latter systems, the non-linearity is obtained by simultaneously operating close to the narrow paramagnetic resonance (at microwave frequencies) and electronic resonance (at optical frequencies) of the rare earth dopant. If the chosen dopant is erbium, with an optical resonance near 1.5~$\mu$m, this approach has the additional advantage of converting to the low-loss fiber telecommunications band. Only a low efficiency has been demonstrated so far, of 10$^{-5}$ in 0.001\% Er:\yso at 5~K  with $\approx 1\,$MHz bandwidth \cite{Fernandez2017}.

A high nonlinearity, and thus conversion efficiency, in these rare earth systems requires a high rare earth concentration and low optical and spin transition linewidths \cite{Williamson2014} so the system can be operated as close to the resonances as possible. However, these two desires are in conflict in dilute rare earth crystals. The rare earth dopant causes strain in the lattice, and as the concentration increases this leads to inhomogeneous broadening of the spectral line \cite{Sellars2004}.  Fully concentrated crystals, however, can also show narrow optical lines along with high optical depth, because the disorder due to randomly distributed dopant ions is removed. For example, a linewidth of 25~MHz has been observed in \isoeuform \cite{Ahlefeldt2016}, comparable to the narrowest linewidths seen in dilute crystals \cite{Thiel2011, Macfarlane1998}.

Fully concentrated crystals also offer another feature: at temperatures around 1~K, they can display magnetic order. The same magnon modes used in ferromagnetic YIG for up-conversion are expected in pure rare earth crystals, although the low ordering temperature means these have not been well studied. YIG does, however, provide a striking example that narrow lines are possible on these modes: the magnon resonance linewidth is as low as 0.6~MHz \cite{Spencer1959, Nakamura2014}
a result that first drove interest in using YIG for up-conversion \cite{Nakamura2016, Haigh2015,Haigh2018} . While the magnon lines are narrow, the available optical transitions in YIG come from iron and have linewidths of hundreds of terahertz, which limits the achievable non-linearity. This is not an issue in concentrated rare earth crystals due to their narrow lines.

In this paper, we propose and analyse a system for up-conversion that combines the advantages of the ferromagnetic magnon and rare earth approaches. We suggest using a magnetically ordered crystal fully concentrated in an optically active rare earth ion. These systems are exciting for microwave to optical transduction because they promise very high atomic concentrations, but at the same time narrow optical transitions and narrow collective magnetic resonances.

\section{Device Overview \label{sec:device_overview}}

\begin{figure}[h]
\centering
\includegraphics[width = 0.5\textwidth]{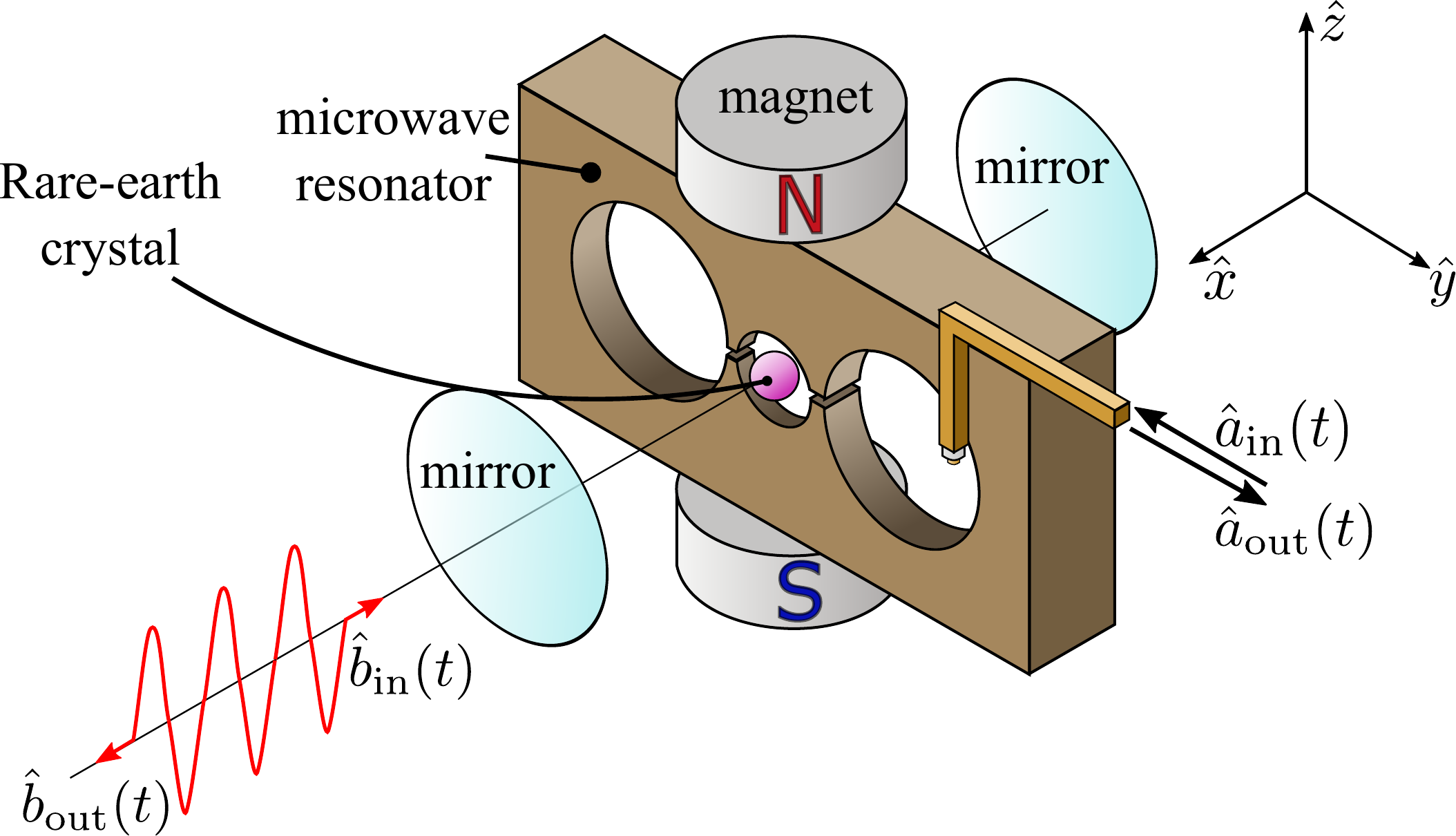}
\caption{The device used to convert microwave photons into optical photons. A rare-earth crystal (doped rare earth crystal for previous device \cite{Williamson2014}, fully concentrated for proposed device) is placed within a microwave resonator and an optical cavity. A static magnetic field is then applied in the $\hat{z}$ direction. This controls the the frequency of the spin resonance (magnon resonance in current device). }
\label{fig:microwave_cavity_containing_sphere}
\end{figure}

\begin{figure}[h!]
\centering
\includegraphics[width = 0.45\textwidth]{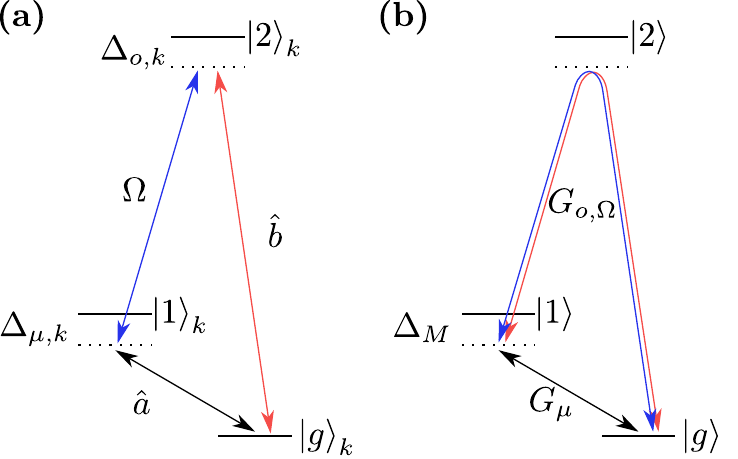}
\caption{(a) Energy level diagram used to describe the dynamics of the previous conversion device \cite{Williamson2014}. An input microwave photon, in mode $\hat{a}$, excites a spin excitation in the $k$'th rare-earth ion, state $\ket{1}_k$. A coherent driving field with Rabi frequency $\Omega$ then drives the spin excitation to an optical excitation, state $\ket{2}_k$. From state $\ket{2}_k$ the device decays into the ground state $\ket{g}_k$ releasing an optical photon in mode $\hat{b}$ in the process. Each state transition is driven off-resonantly indicated by the detunings $\Delta_{o,k}$ and $\Delta_{\mu,k}$. (b) Energy level diagram used to describe the dynamics of our newly proposed conversion device. Strong interactions between the spins means that using the single atom picture becomes problematic. The conversion occurs via a similar off-resonant Raman-like process to that of the previous conversion device. The state $\ket{1}$ has one excitation in the Kittel mode and is driven off-resonance (an amount $\Delta_M$) by both a microwave input field (coupling strength $G_{\mu}$) and a two optical photon Raman process (coupling strength $G_{o,\Omega}$). }
\label{fig:system_energy_diagram}
\end{figure}

Our proposed device uses a similar process to a device proposed for low-concentration doped rare earth crystals \cite{Williamson2014, Fernandez2015, Fernandez2017}, and we first briefly describe that device. The device is shown  in Fig.~\ref{fig:microwave_cavity_containing_sphere}. A doped rare-earth ion crystal at cryogenic temperatures is coupled to a microwave resonator, an optical resonator and a coherent optical driving field. The dynamics of the device can be described by an off-resonant Raman-like process using the three-level energy diagram shown in Fig.~\ref{fig:system_energy_diagram} (a). The rare-earth ions begin in their ground state $\ket{g}$, a microwave photon (frequency $\omega_{\mu}$) input into the microwave cavity excites a spin excitation within the doped rare-earth crystal, state $\ket{1}$. A coherent driving field (frequency $\omega_{\Omega}$) then excites an optical transition, state $\ket{2}$, where the rare-earth ion then returns to the ground state, emitting an optical photon resonant with the optical cavity (frequency $\omega_o$). All three fields are detuned from their respective transition resonances, but they are kept in three photon resonance, $\omega_{\mu} + \omega_{\Omega} = \omega_{o}$. Driving all transitions off-resonantly greatly simplifies the dynamics of the device, because it allows us to adiabatically eliminate the dynamics of the excited levels.

Our newly proposed device works in the same manner as the device just described, however, in the new device we replace the doped rare-earth ion crystal with a fully concentrated rare-earth ion crystal. In a high concentration regime the strong interactions between rare-earth ions means the dynamics are best understood in  collective excitations,  rather than the individual ion energy states of the previous device. Similar to the previous device, the dynamics of the newly proposed device can be described by an off-resonant Raman-like process using the three level energy diagram shown in Fig.\ref{fig:system_energy_diagram} (b), however now state $\ket{1}$ is a single collective spin excitation (a magnon) that is excited both by direct microwave driving and by a two-optical-photon Raman process.

In our device we will be working in the regime where the microwave field wavelength is much greater than the rare-earth ion crystal sample size. Under this condition the magnon modes that can be excited within the sample do not propagate, and are known as magnetostatic modes. These modes are well understood in an isotropic spherical material \cite{Walker1957, Fletcher1959, Roschmann1977}. The description of our newly proposed conversion device assumes that an input microwave photon excites only the lowest order, spatially uniform $(1,1,0)$ magnon mode or Kittel mode. This assumption is justified by operating with the other magnon modes highly detuned from both the microwave driving field and the driving by the two optical fields. Furthermore, choosing a microwave resonator with a uniform mode across the sample will mean that the other magnetostatic modes are only weakly driven, since those modes are spatially non-uniform.

We derive the device Hamiltonian in Appendix \ref{sec:hamiltonian}, and show that it can be reduced to a coupling between the microwave ($\Hat{a}$) and optical ($\hat{b}$) modes

\begin{equation}
    \mathcal{H} = \hbar(\xi \Hat{a}^\dagger\Hat{b} + \xi^*\Hat{b}^{\dagger}\Hat{a})
\end{equation}

where the coupling strength, $\xi$, is given by

\begin{equation}
    \xi = \frac{G_{\mu}^*G_{o,\Omega}}{\Delta_M}.
\end{equation}
with $\Delta_M$ the microwave detuning from the Kittel mode. The strength of the Kittel mode coupling via the the microwave input field ($G_{\mu}$) and the two optical photon Raman process ($G_{o,\Omega}$) are derived in Appendix \ref{sec:hamiltonian} to be

\begin{align}
    G_{\mu} &= \frac{1}{\sqrt{N}}\sum_kg_{\mu,k}  \label{eq:couplingfactors1}\\
    G_{o,\Omega} &= \frac{1}{\sqrt{N}}\sum_k \frac{\Omega_k^*g_{o,k}}{\Delta_{o,k}}
    \label{eq:couplingfactors2}
\end{align}

Here, the sum runs over all $N$ rare-earth ions in the sample, $g_{\mu,k}$ and $g_{o,k}$ represent the coupling strength between the $k$'th ion and the microwave and optical cavities respectively, $\Omega_k$ is the Rabi frequency of the classical driving field and $\Delta_{o,k}$ is the detuning of the optical field from the $k$'th ions optical transition.


\section{Device efficiency\label{sec:deveff}}

In this section the input-output formalism developed in  \cite{Collett1984} is used to calculate the relations between the microwave and optical cavity fields and their respective input/output modes. This allows us to define an impedance matching condition for the device, which we re-express in terms of interpretable parameters.

Applying the equation of motion for the field in a one-sided cavity, given by Eq.(6) and Eq.(19) in \cite{Collett1984}, the two cavity fields in the device evolve as:

\begin{equation}
\begin{split}
    \dot{a}(t) &= -i\xi^*\hat{b}(t) - \frac{\kappa_{\mu}}{2}\hat{a}(t) + \sqrt{\kappa_{\mu}}\hat{a}_{\text{in}}(t) \\
    \dot{b}(t) &= -i\xi\hat{a}(t) - \frac{\kappa_o}{2}\hat{b}(t) + \sqrt{\kappa_o}\hat{b}_{\text{in}}(t)
\end{split}
\label{eq:cavity_eqn_motion}
\end{equation}

where $\kappa_{\mu}$ and $\kappa_o$ represent the decay rates of the two cavities. Applying Eq.(5) in \cite{Collett1984} also gives

\begin{equation}
\begin{split}
    \dot{a}_{out}(t) + \dot{a}_{in}(t) &= \sqrt{\kappa_\mu}\dot{a}(t)\\
    \dot{b}_{out}(t) + \dot{b}_{in}(t) &= \sqrt{\kappa_o}\dot{b}(t)
\end{split}
\label{eq:in_out_relation}
\end{equation}

Fourier transforming Eq.\eqref{eq:cavity_eqn_motion} and using Eq.\eqref{eq:in_out_relation} gives

\begin{equation}
\begin{split}
    \hat{a}_{\text{out}}(\omega) = &-\frac{|\xi|^2 - \left(\frac{\kappa_o}{2} - i\omega\right)\left(\frac{\kappa_{\mu}}{2} + i\omega\right)}{|\xi|^2 + \left(\frac{\kappa_o}{2} + i\omega\right)\left(\frac{\kappa_{\mu}}{2} + i\omega\right)}\hat{a}_{\text{in}}(\omega) \\
    &-\frac{i\xi^*\sqrt{\kappa_{\mu}\kappa_o}}{|\xi|^2 + \left(\frac{\kappa_o}{2} + i\omega\right)\left(\frac{\kappa_{\mu}}{2} + i\omega\right)}\hat{b}_{\text{in}}(\omega)
    \label{eq:input_output_1}
\end{split}
\end{equation}

\begin{equation}
\begin{split}
    \hat{b}_{\text{out}}(\omega) = &-\frac{|\xi|^2 - (\frac{\kappa_o}{2} - i\omega)(\frac{\kappa_{\mu}}{2} + i\omega)}{|\xi|^2 + (\frac{\kappa_o}{2} + i\omega)(\frac{\kappa_{\mu}}{2} + i\omega)}\hat{b}_{\text{in}}(\omega) \\
    &-\frac{i\xi\sqrt{\kappa_{\mu}\kappa_o}}{|\xi|^2 + (\frac{\kappa_o}{2} + i\omega)(\frac{\kappa_{\mu}}{2} + i\omega)}\hat{a}_{\text{in}}(\omega)
    \label{eq:input_output_2}
\end{split}
\end{equation}

Where $\omega$ is the detuning from both the microwave and optical cavity resonances. For both Eq.\eqref{eq:input_output_1} and Eq.\eqref{eq:input_output_2}, the first terms on the right side describe the signals reflected from the cavities. The second terms give the photon conversion between the microwave and optical fields. From these expressions the number conversion efficiency is:

\begin{equation}
    \eta(\omega) = \left|\frac{\xi\sqrt{\kappa_{\mu}\kappa_o}}{|\xi|^2 + (\frac{\kappa_o}{2} + i\omega)(\frac{\kappa_{\mu}}{2} + i\omega)}\right|^2
\end{equation}

Provided the microwave and optical fields are resonant with the cavities ($\omega\ll \kappa_o/2$, $\omega\ll \kappa_\mu/2$)  and the strength of the coupling between the microwave and optical fields is chosen appropriately ($2|\xi|=\sqrt{\kappa_{\mu}\kappa_o}$) to achieve impedance matching, an input microwave field is completely converted into an output optical field and \emph{vice versa}. The impedance matching condition can also be written as
\begin{align}
    1 = \frac{2|G_{\mu}||G_{o,\Omega}|}{\kappa\Delta_M}\,.
    \label{eq:impedance_matching}
\end{align}
The bandwidth of the device, meanwhile, is determined by the smaller of $\kappa_o$ and $\kappa_\mu$. For our feasibility analysis later we will assume that they are equal to $\kappa_o=\kappa_\mu=\kappa$, giving a conversion bandwidth of $\sqrt{2}\kappa$.

To investigate if the impedance condition can realistically be achieved, we derive simplified expressions for $G_\mu$, $G_{o,\Omega}$. Setting the dipole moments $\bm{\mu}_{g1,k}$, $\bm{d}_{g2,k}$, $\bm{d}_{12,k}$, introduced in Eq. \eqref{eq:H_I}, as homogeneous across the sample (independent of $k$), we can then take their scalar projections along the microwave and optical mode functions $\bm{\chi}(\bm{r})$, $\bm{\phi}(\bm{r})$ and $\bm{\epsilon}(\bm{r})$, introduced in Eqs.\eqref{eq:B_mu}-\eqref{eq:E_omeg}. Because the optical detuning is much larger than the optical inhomogeneous linewidth we also take $\Delta_{o,k}$ to be the same for each atom.
Thus our expression for $G_{\mu}$ and $G_{o,\Omega}$ become,

\begin{align}
    G_{\mu}  &= \frac{1}{\sqrt{N}}\sqrt{\frac{\omega_{\mu}\mu_0}{2\hbar V_{\mu}}}\mu_{g1}\sum_{k=1}^N|\bm{\chi}(\bm{r}_k)|\\
    G_{o,\Omega} &= \frac{1}{\sqrt{N}}\sqrt{\frac{\omega_o}{2\hbar\epsilon_0V_o}}\frac{d_{g2}\Omega_0}{\Delta_o}\sum_{k=1}^N |\bm{\phi}(\bm{r}_k)||\bm{\epsilon}(\bm{r}_k)|
\end{align}

Approximating the sums as integrals over the crystal volume ($V_c$) with atomic density $\rho=N/V_c$, we get

\begin{align}
    G_{\mu}  &= \sqrt{N}\sqrt{\frac{\omega_{\mu}\mu_0}{2\hbar V_{\mu}}}\mu_{g1}\frac{1}{V_c}\int_{V_c} d^3r \ |\bm{\chi}(\bm{r}_k)| \nonumber \\
    &= \sqrt{\rho V_c}\sqrt{\frac{\omega_{\mu}\mu_0}{2\hbar V_{\mu}}}\mu_{g1} \label{eq:G_mu_simplified}\\
    G_{o,\Omega} &= \sqrt{N}\sqrt{\frac{\omega_o}{2\hbar\epsilon_0V_o}}\frac{d_{g2}\Omega_0}{\Delta_o}\frac{1}{V_c}\int_{V_c} d^3r \ |\bm{\phi}(\bm{r}_k)||\bm{\epsilon}(\bm{r}_k)| \nonumber \\
    &= \sqrt{\rho V_c}\sqrt{\frac{\omega_o}{2\hbar\epsilon_0V_o}}\frac{d_{g2}\Omega_0}{\Delta_o} F \label{eq:G_o_simplified}
\end{align}
Where we have defined $F$ as the integral over the optical fields, and the integral over the microwave field is simply $V_c$ because we assumed the field was uniform over the sample in Section \ref{sec:device_overview}.

\section{Feasibility}
In this section, we show that achieving unit efficiency is feasible in the  device we propose using existing materials and reasonable experimental conditions. We assume the conversion operates between a 5\,GHz frequency microwave photon (similar to the frequency used for superconducting qubits), and an optical photon in the 1550~nm telecommunications band.
We consider a 2~mm diameter \erform sphere and a loop-gap and Fabry P\'{e}rot resonator.

\begin{table*}
\setlength{\tabcolsep}{15pt}
\renewcommand{\arraystretch}{1.2}

\begin{tabular}{lcr}
	\hline
	\vspace{-3mm}\\
	Parameter & Symbol & Value \\
	\vspace{-3mm}\\
	\hline
	\hline
	\vspace{-3mm}\\

	Crystal volume & $V_c$ & $\SI{4.2e-9}{\meter\cubed}$ \\

	Atomic number density & $\rho$ & $\SI{4.0e27}{\per\meter\cubed}$ \cite{Kepert1983}\\

	Microwave angular frequency & $\omega_\mu$ & $2\pi\times\SI{5}{\giga\hertz}$ \\

	Optical angular frequency & $\omega_o$ & $2\pi\times\SI{195}{\tera\hertz}$\\

	Maximum available magnetic dipole moment, $\ket{g}\to\ket{1}$ & $\max(\mu_{g1})$ & \SI{3.0e-23}{\joule\per\tesla} \cite{Couture1984}\\

  Electric dipole moment, $\ket{g}\to\ket{2}$ & $d_{g2}$ & \SI{2.0e-32}{\coulomb\meter} \footnote{Measured from experiments on Er:\euform, which is isostructural to \erform.}\\

	Optical transition FWHM & $\gamma_o$ & \SI{1.24}{\giga\hertz} \footnote{Measured from absorption spectroscopy experiments performed on \erform at 5 K and 3.5 T, unpublished.}\\

	Microwave cavity mode volume & $V_\mu$ & $\SI{2.9e-7}{\meter\cubed}$ \footnote{Obtained from the numerical model of the loop gap resonator. See supplementary material.}\\

	Optical cavity mode volume & $V_o$ & $\SI{2.9e-11}{\meter\cubed}$ \footnote{Obtained from the numerical model of the Fabry P\'{e}rot resonator. See supplementary material.}\\

	Optical field overlap integral & $F$ & $\SI{2.4e-4}{}$ \footnotemark[4]\\

	Maximum available Rabi frequency & $\max(\Omega_0$) & \SI{68}{\mega\hertz} \footnote{Calculated assuming a maximum pump laser power of $\SI{1}{\micro\watt}$, a typical tolerable load for cryogenic systems. The beam width and cavity field enhancement was calculated using the Fabry P\'{e}rot resonator model. See supplementary material.}\\

	Maximum available microwave cavity quality factor & $\max(Q_\mu)$ & \SI{9e4}{} \footnote{Easily achieved with 3D copper microwave resonators \cite{Fernandez2015}}\\

	Maximum available optical cavity quality factor & $\max(Q_o)$ & \SI{3e8}{} \footnote{Measured in a whispering gallery mode optical resonator made from a doped rare earth ion crystal.\cite{inprep}}\\

	\vspace{-3mm}\\
	\hline
\end{tabular}
\caption{Table of parameters for the proposed device for a 2~mm diameter \erform crystal,  loop-gap microwave resonator and Fabry P\'{e}rot optical resonator. When the parameter can take on a range of values, the relevant extremum is given.}
\label{tab:parameter_values}
\end{table*}

The rare earth crystal that we consider is erbium chloride hexahydrate (\erform), a crystal that orders ferromagnetically below 350~mK \cite{Lagendijk1973}. The optical transition of Er between the $^4$I$_{15/2}$ ground state and the $^4$I$_{13/2}$ excited state occurs at 1540 nm. A modest applied field would bring the frequency of the Kittel mode to 5\,GHz.


\begin{figure}[h!]
\centering
\includegraphics[width = 0.4\textwidth]{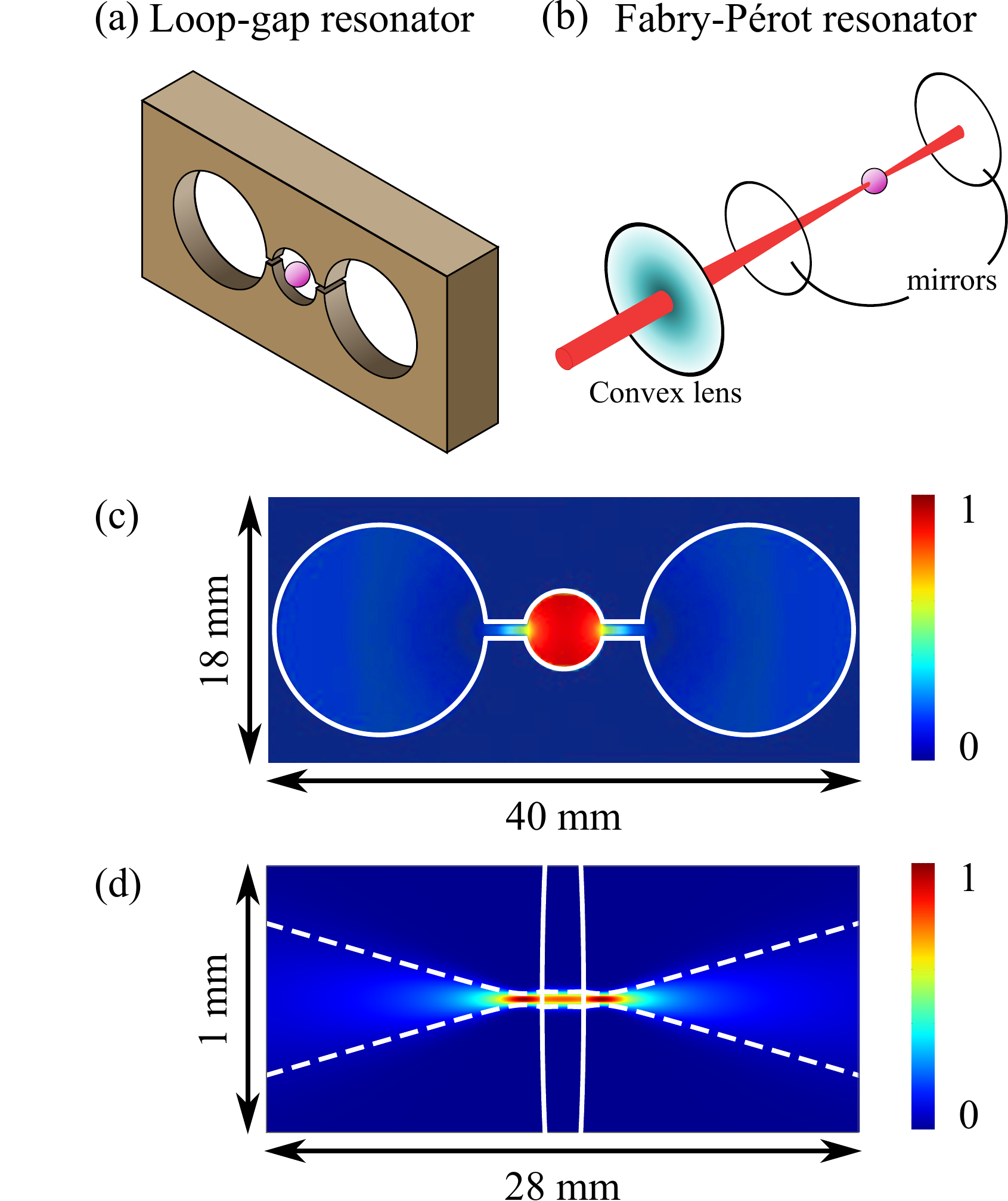}
\caption{Diagram showing proposed geometry (not to scale) for our (a) microwave resonator and (b) optical resonator. A convex lens is used in the optical resonator to focus the beam into the center of the sphere in order to minimize spherical aberration. (c, d) show the modeled field amplitudes of the microwave and optical resonators. The solid white lines in (c) outline the loop-gap, and in (d) outline the crystal of diameter 2 mm. The dashed line in (d) indicate the gaussian beam width.}
\label{fig:resonator_design}
\end{figure}

The optical and microwave cavities were modeled to determine the experimental parameters possible. The models are shown in Figure \ref{fig:resonator_design}. The microwave cavity is a shielded loop-gap resonator, which produces a highly uniform magnetic field in the center of the cavity. We used finite-difference time domain solutions to model the loop-gap resonator for a 5\,GHz resonance. The optical cavity was a Fabry-P\'{e}rot resonator with curved mirrors and had a 27\,$\mu$m waist inside the middle of the spherical sample. We use ray tracing with the paraxial approximation to model a Fabry P\'{e}rot resonator with a 1540 nm resonance and 5\,GHz free spectral range. The 5\,GHz free spectral range was chosen so that both optical fields are resonant with the cavity.

To justify the feasibility of the conversion device, we need to show four things: (a) that the impedance matching condition, Eq.\eqref{eq:impedance_matching}, can be satisfied with physically realistic values, (b)  that the Rabi frequency and microwave cavity couplings are small compared to the optical detuning ($\Omega_0,g_\mu\ll\Delta_o$), (c) that the cavity coupling to the magnon mode is small compared to the magnon detuning ($G_\mu, G_{o,\Omega}\ll \Delta_M$), and (d) that the optical transition linewidth is small compared to the optical detuning ($\gamma_o\ll\Delta_o$). Satisfying these four criteria will guarantee unit conversion efficiency, however it is also desirable to maximise the device conversion bandwidth, $\sqrt{2}\kappa$. This introduces a fifth design criteria (e), when possible, maximise the device bandwidth $\sqrt{2}\kappa\propto G_\mu G_{o,\Omega}$ (see Eq.\eqref{eq:impedance_matching}).

We first consider satisfying the off-resonant condition (d). From Table~\ref{tab:parameter_values}, the optical linewidth is $\gamma_o=\SI{1.24}{\giga\hertz}$. By enforcing that the optical detuning is five times larger than the linewidth, $\Delta_o=\SI{6.2}{\giga\hertz}$, we satisfy condition (d).

We now consider adiabatic condition (b). From Table~\ref{tab:parameter_values}, the maximum Rabi frequency possible for the proposed device is $\Omega_0=\SI{68}{\mega\hertz}$, assuming a maximum pump laser power of \SI{1}{\micro\watt}. This is much less than the \SI{6.2}{\giga\hertz} optical detuning, so $\Omega_0\ll\Delta_0$. From the parameters in Table~\ref{tab:parameter_values}, the coupling between the microwave cavity and individual ions $g_\mu$, given by Eq.\eqref{eq:g_mu}, is between 0 and \SI{0.8}{\hertz}, depending on the direction of the microwave field relative to the anisotropic transition dipole moment $\mu_{g1}$. Thus $g_\mu\ll\Delta_0$, so condition (b) is satisfied.

We now consider adiabatic condition (c) and design parameter (e). From the parameters given in Table~\ref{tab:parameter_values}, the coupling strength between the  microwave cavity and the magnon mode $G_\mu$, given by Eq.\eqref{eq:G_mu_simplified}, is  between 0 and 3.1~GHz, depending on the direction of the microwave field relative to the anisotropic transition dipole moment $\mu_{g1}$. Meanwhile, the coupling between the optical cavity and the magnon mode $G_{o,\Omega}$ (Eq.\eqref{eq:G_o_simplified}) is between 0 and 63~MHz, depending on the choice of the Rabi frequency $\Omega_0$. The largest feasible detuning from the Kittel mode $\Delta_M$, is given by the frequency spacing to the next magnon mode. At 5\,GHz this is expected to be $\mathcal{O}(\SI{100}{\mega\hertz})$ \cite{Fletcher1959}. Given the tunability of $G_\mu,G_{o,\Omega}$, we can satisfy (e) by maximising them within the constraint of condition (c). This can be achieved by selecting $G_\mu=G_{o,\Omega}=\SI{10}{\mega\hertz}$ and $\Delta_M=\SI{100}{\mega\hertz}$, where the factor of ten difference guarantees $G_\mu,G_{o,\Omega}\ll\Delta_M$.

Finally, we consider the impedance matching condition (a). To satisfy Eq.~\eqref{eq:impedance_matching} with the values of $G_\mu,G_{o,\omega},\Delta_M$ chosen above requires the cavity decay rates to be $\kappa=\SI{2}{\mega\hertz}$. This necessitates the quality factors of the microwave and optical cavities to be $Q_\mu=2.5\times10^3$ and $Q_o=9.7\times10^7$ respectively. From Table~\ref{tab:parameter_values}, this is within the range of feasible quality factors, so satisfying (a) is physically achievable.

This means our proposed device can satisfy all the requirements of the theory set out in Appendix ~\ref{sec:hamiltonian} and Section ~\ref{sec:deveff} with bandwidths well in excess of one megahertz possible. It should be noted that when choosing parameters the limiting factor on the bandwidth was the spacing between the magnon modes. The bandwidth available is set by the collective coupling rates which were in turn limited by the adiabatic condition to be less than the magnon detuning, which is in turn limited by the spacing between the magnon modes. This means that the reliability of this feasibility analysis is somewhat insensitive to all the other system parameters. It is unfortunate that of all the parameters we have estimated, the spacing between the magnon modes is the least well constrained, in particular the calculations in \cite{Fletcher1959} are for an isotropic ferromagnet, which \erform is not. For the optical properties of \erform and the Zeeman sensitivity of the spin states we used measured quantities, but collective magnetic resonance is poorly studied in crystals with rare earth ions as the only magnetic ion. The only observation we are aware of is Ref. \onlinecite{Abraham1992}. This is perhaps because rare earth ions have weak exchange interactions and only order at temperatures comparable to 1\,K. Additionally, narrow collective resonances are only seen at temperatures well below the Curie temperature \cite{Vonsovskii1966}  and this requirement for very low temperatures would rule such materials out for many traditional applications of magnetic materials. This is not a problem when using these materials for frequency conversion of microwave photons because  microwave frequency quantum states, already require these very low temperatures.

\section{Conclusion}

We proposed using insulating crystals fully concentrated with rare earth ions for microwave-to-optical frequency conversion. These materials promise very narrow optical transitions at the same time as very high ion densities and narrow collective magnetic resonances. The analysis presented showed that using \erform as the crystal, high efficiency, high bandwidth microwave-to-optical conversion should be possible. This result suggests further study of fully concentrated rare earth crystals, particularly their collective resonances, would be a promising avenue.

\section{Acknowledgements}
The authors would like that M. Sellars for useful discussions. JRE and JJL were supported by the Marsden Fund of the Royal Society of New Zealand through Contract No.\ UOO1520. RLA is a recipient of an Australian Research Council Discovery Early Career Researcher Award (project No. DE170100099) funded by the Australian Government.

\appendix

\section{Device Hamiltonian}\label{sec:hamiltonian}

The Hamiltonian for our device can be written as

\begin{equation}
\begin{split}
    \mathcal{H} &= H_{\text{F}} + H_{\text{A}} + H_{\text{IF}} \\
    &+ H_{\text{E}} + H_{\text{D}}.
\end{split}
\end{equation}

Here, $H_{\text{F}}$ describes the energy in the cavity fields
\begin{equation}
    H_{\text{F}} = \hbar\omega_{\mu,c}\hat{a}^{\dagger}\hat{a} + \hbar\omega_{o,c}\hat{b}^{\dagger}\hat{b},
\end{equation}

where $\hat{a}$ and $\hat{b}$ are the annihilation operators of the microwave and optical cavities respectively and $\omega_{\mu,c}$ and $\omega_{o,c}$ are the (bare) resonant frequencies of the microwave and optical cavities respectively.

\vspace{5mm}

$H_{\text{A}}$ describes the energy of the rare-earth ions

\begin{equation}
    H_{\text{A}} =  \sum_k \hbar\omega_{2,k}\sigma_{22,k} + \hbar\omega_{1,k}\sigma_{11,k}.
\end{equation}

The sum here is over all rare-earth ions within the sphere and the subscript $k$ indicates  the $k$th rare-earth ion. The resonant frequency of the excited states $\ket{1}_k$ and $\ket{2}_k$ are given by $\omega_{1,k}$ and $\omega_{2,k}$ respectively, and $\sigma_{ij} \equiv \ket{i}\bra{j}$ represents the atomic transition operator. This term has taken into account the energy splitting due to the Zeeman interaction, such that $\hbar\omega_{1,k} = g\mu_B B_0 $, where $g$ is the Land\'e g-factor, $\mu_B$ is the Bohr magneton and $B_0$ is the strength of the static magnetic field.

\vspace{5mm}

$H_{\text{IF}}$ describes the interaction between the rare-earth ions and the cavity fields

\begin{equation}
    H_{\text{IF}} = \sum_k \bm{\mu}_{g1,k}\cdot\bm{B}_{\mu}(\bm{r}_k) + \bm{d}_{g2,k}\cdot\bm{E}_o(\bm{r}_k) + \bm{d}_{12,k}\cdot\bm{E}_{\Omega}(\bm{r}_k,t). \label{eq:H_I}
\end{equation}

Here, $\bm{\mu}_{g1}$ is the magnetic dipole operator of the microwave transition and $\bm{d}_{12}$ and $\bm{d}_{g2}$ are the electric dipole operators of the optical and driving field transitions respectively.  $\bm{B}_{\mu}, \bm{E}_o$ and $\bm{E}_{\Omega}$ describe the magnetic and electric field operators for the microwave, optical and driving fields respectively,

\begin{align}
    \bm{B}_{\mu} &= \sqrt{\frac{\hbar \omega_{\mu,c}}{2\epsilon_0V_{\mu}}}(\Hat{a}^{\dagger} + \Hat{a})\bm{\chi}(\bm{r}_k) \label{eq:B_mu}\\
    \bm{E}_{o} &= \sqrt{\frac{\hbar \omega_{o,c}}{2\epsilon_0V_o}}(\Hat{a}^{\dagger} + \Hat{a})\bm{\phi}(\bm{r}_k) \\
    \bm{E}_{\Omega} &= E_0(e^{-i\omega_{\Omega}t} + e^{i\omega_{\Omega}t})\bm{\epsilon}(\bm{r}_k). \label{eq:E_omeg}
\end{align}

where the mode volumes of the microwave and optical cavities are represented by $V_{\mu}$ and $V_o$ respectively and $E_0$ is the peak magnitude of the coherent driving field. The microwave and optical mode functions are represented by $\bm{\chi}(\bm{r})$, $\bm{\phi}(\bm{r})$ and $\bm{\epsilon}(\bm{r})$ respectively, and have been normalized between 0 and 1.

\vspace{5mm}

Expanding Eq.\eqref{eq:H_I} out we can obtain the expression

\begin{equation}
\begin{split}
    H_{\text{IF}} = \sum_k &\hbar g_{\mu,k}(\sigma_{g1,k} + \sigma_{1g,k})(\hat{a}^{\dagger} + \hat{a}) \\
    + &\hbar g_{o,k}(\sigma_{g2,k} + \sigma_{2g,k})(\hat{b}^{\dagger} + \hat{b}) \\
    + &\hbar\Omega_{k} (\sigma_{12,k} + \sigma_{21,k})(e^{-i\omega_{\Omega}t} + e^{i\omega_{\Omega}t}).
\end{split}
\end{equation}

Here the coupling strengths between the ions and the microwave and optical cavities are given by $g_{\mu}$ and $g_{o}$ respectively,

\begin{align}
    g_{\mu,k} &= \sqrt{\frac{\omega_{\mu}\mu_0}{2\hbar V_{\mu}}}\bm{\mu}_{g1,k}\cdot\bm{\chi}(\bm{r}_k) \label{eq:g_mu}\\
    g_{o,k} &= \sqrt{\frac{\omega_o}{2\hbar\epsilon_0V_o}}\bm{d}_{g2,k}\cdot\bm{\phi}(\bm{r}_k). \label{eq:g_o}
\end{align}

and the Rabi frequency of the classical driving field for each ion is represented by $\Omega$,

\begin{equation}
\begin{split}
    \Omega_k & = \frac{1}{\hbar}\bm{d}_{12,k}\cdot E_0\bm{\epsilon}(\bm{r}_k) \label{eq:Omega_k} \\
    &= \bm{\Omega}_{0,k}\cdot \bm{\epsilon}(\bm{r}_k)
\end{split}
\end{equation}

where $\bm{\Omega}_{0,k}$ is the peak Rabi frequency.

\vspace{5mm}

The final two terms describe interactions between the spins that generate collective behavior. $H_{\text{E}}$ describes the exchange interaction between neighboring spins,

\begin{equation}
    H_{\text{E}} = - J\sum_{k,\delta}\hat{\bm{S}}_k\cdot\hat{\bm{S}}_{k + \delta}
\end{equation}

Here, the $\delta$ summation runs over all nearest neighbours to the $k$th rare earth ion and $J$ is the isotropic exchange constant.

\vspace{5mm}

Finally, $H_{\text{D}}$ describes the dipole-dipole interaction between spins,

\begin{equation}
    H_{\text{D}} = \frac{1}{2}g^2\mu_B^2\sum_{k,j}\left[\frac{\hat{\bm{S}}_k\cdot\hat{\bm{S}}_j}{r_{kj}^3} - \frac{3(\hat{\bm{S}}_k\cdot\bm{r}_{kj})(\hat{\bm{S}}_j\cdot\bm{r}_{kj})}{r_{kj}^5} \right]
\end{equation}

 Here, both the $k$ and $j$ summations run over all rare-earth ions, and $\bm{r}_{kj}$ is the displacement vector from the $k$th rare-earth ion to the $j$th rare-earth ion.

\vspace{5mm}

As we are working within a regime where only magnetostatic modes are excited, the effect of the exchange interaction is negligible and can be ignored \cite{Walker1957,Fletcher1959}. Further,  following the method of Holstein and Primakoff \cite{White2007}, it can be shown that for a spherical sample in which only the Kittel mode is excited, the dipole-dipole interaction reduces to a constant which can be ignored when analysing the dynamics of the device.

\vspace{5mm}

Before moving our Hamiltonian into the interaction picture we rewrite $H_{\text{F}}$ and $H_{\text{A}}$ in the following way

\begin{equation}
\begin{split}
    H_{\text{F}} &= \hbar\omega_o\hat{b}^{\dagger}\hat{b} + \hbar\omega_{\mu}\hat{a}^{\dagger}\hat{a} + \hbar(\omega_{\mu,c} - \omega_{\mu})\hat{a}^{\dagger}\hat{a} + \hbar(\omega_{o,c} - \omega_o)\hat{b}^{\dagger}\hat{b} \\
    &= \hbar\omega_o\hat{b}^{\dagger}\hat{b} + \hbar\omega_{\mu}\hat{a}^{\dagger}\hat{a} + \hbar\delta_{\mu}\hat{a}^{\dagger}\hat{a} + \hbar\delta_o\hat{b}^{\dagger}\hat{b}
\end{split}
\end{equation}

\begin{equation}
\begin{split}
    H_{\text{A}} = \sum_k &\hbar\omega_{o}\sigma_{22,k} + \hbar\omega_{\mu}\sigma_{11,k} \\
    +&\hbar(\omega_{2,k} - \omega_o)\sigma_{22,k} + \hbar(\omega_{1,k} - \omega_{\mu})\sigma_{11,k} \\
    = \sum_k &\hbar\omega_{o}\sigma_{22,k} + \hbar\omega_{\mu}\sigma_{11,k} \\
    +&\hbar\Delta_{o,k}\sigma_{22,k} + \hbar\Delta_{\mu,k}\sigma_{11,k}
\end{split}
\end{equation}

Here, $\omega_o$ and $\omega_{\mu}$ are the frequencies of the applied optical and microwave fields respectively. The detunings of the microwave and optical fields from the cavities are $\delta_{\mu}$ and $\delta_o$ respectively and the detuning of the microwave and optical fields from the spin and optical transitions are $\Delta_{\mu}$ and $\Delta_o$ respectively.

\vspace{5mm}

Writing our Hamiltonian as $\mathcal{H} = H_0 + V$, where

\begin{equation}
\begin{split}
    H_0 &= \hbar\omega_{\mu}\Hat{a}^{\dagger}\Hat{a} + \hbar\omega_{o}\Hat{b}^{\dagger}\Hat{b} \\
    &+ \sum_k(\hbar\omega_{o}\sigma_{22,k} + \hbar\omega_{\mu}\sigma_{11,k})
\end{split}
\end{equation}

\begin{equation}
\begin{split}
    V &= \hbar\delta_{\mu}\hat{a}^{\dagger}\hat{a} + \hbar\delta_o\hat{b}^{\dagger}\hat{b} \\
    &+ \sum_k(\hbar\Delta_{o,k}\sigma_{22,k} + \hbar\Delta_{\mu,k}\sigma_{11,k}) + H_{\text{I}}.
\end{split}
\end{equation}


we move into the interaction picture via the transformation $\mathcal{H} = e^{iH_0t/\hbar}Ve^{-iH_0t/\hbar}$ to get

\begin{equation}
\begin{split}
    \mathcal{H} &= \hbar\delta_{\mu}\Hat{a}^{\dagger}\Hat{a} + \hbar\delta_o\hat{b}^{\dagger}\hat{b} \\
    &+ \sum_{k}(\hbar\Delta_{o,k}\sigma_{22,k} + \hbar\Delta_{\mu,k}\sigma_{11,k}) + \sum_{k}(\Omega_k\sigma_{21,k} + \text{H.c.}) \\
    &+ \sum_{k}(g_{\mu,k}\sigma_{1g,k}\hat{a} + \text{H.c.}) + \sum_{k}(g_{o,k}\sigma_{2g,k}\hat{b} + \text{H.c.})
\label{eq:H0}
\end{split}
\end{equation}

We will now simplify our device Hamiltonian by removing the dynamics of the optical and Kittel modes, which will enable the conversion efficiency of our device to be calculated in Section \ref{sec:deveff}. This simplification is possible as our system is driven off-resonantly, thus the $\ket{1}$ and $\ket{2}$ states will remain nearly unpopulated, acting only as intermediate states to enable a three-photon process. Removing the dynamics of the intermediate states can be achieved by the process of adiabatic elimination; for details on the process see \cite{Cohen-Tannoudji1992, Brion2007}.

\vspace{5mm}

We first adiabatically eliminate the optically excited state from each atom. Working with large optical detunings, $|\Delta_{o,k}| \gg |g_{o,k}|$ and $|\Delta_{o,k}| \gg |\Omega_k|$, we can adiabatically eliminate the optically excited states from each atom to obtain,

\begin{equation}
\begin{split}
    \mathcal{H} = \hbar&\delta_{\mu}\Hat{a}^{\dagger}\Hat{a} + \hbar\delta_o\hat{b}^{\dagger}\hat{b} + \hbar\sum_k \Delta_{\mu,k}\sigma_{11,k}\\
    -\hbar&\sum_k \left( \frac{|\Omega_k|^2}{\Delta_{o,k}}\right)\sigma_{11,k} \\ -\hbar&\sum_k\frac{1}{\Delta_{o,k}}|g_{o,k}|^2\sigma_{gg,k}\hat{b}^\dagger\hat{b} \\
    + \hbar&\sum_k\left(g_{\mu,k}\sigma_{1g,k}\hat{a} + \text{H.c.}\right) \\ -\hbar&\sum_k\frac{1}{\Delta_{o,k}}(g_{o,k}\Omega^*_k\sigma_{1g,k}\hat{b} + \text{H.c.}).
    \label{eq:adiab_elim_1}
\end{split}
\end{equation}

Three new terms appear in the Hamiltonian. The fourth term of Eq.\eqref{eq:adiab_elim_1} describes the AC stark shift, which under the adiabatic elimination condition ($|\Delta_{o,k}| \gg |\Omega_k|$) will be small and hence ignored.  The fifth term in Eq.\eqref{eq:adiab_elim_1} represents a shift in the resonance frequency of the optical cavity due to the presence of the atoms. This term as can be compensated for by tuning the resonant frequency of the optical cavity. This can be seen from Eq.\eqref{eq:adiab_elim_1} by rewriting the optical cavity detuning as $\delta_o = \delta_{o1} + \delta_{o2}$ where $\delta_{o1} = \hbar\sum_{k}\frac{1}{\Delta_{o,k}}|g_{o,k}|^2\sigma_{gg,k}$. The final term in Eq.\eqref{eq:adiab_elim_1} describes the coupling between the ground and excited states via the two optical fields.

\vspace{5mm}

To adiabatically eliminate the first excited state, the Kittel mode, we need to express the Hamiltonian in terms of Kittel mode creation and annihilation operators. This transformation will be carried out in several steps. First, we can re-express the Hamiltonian in terms of spin operators, via the transformations:

\begin{equation}
\begin{split}
    \frac{1}{2}(\sigma_{gg,k} - \sigma_{11,k}) &\rightarrow \hat{S}_{z,k} \\
    \sigma_{g1,k} &\rightarrow \hat{S}_{+,k} \\
    \sigma_{1g,k} &\rightarrow \hat{S}_{-,k}
    \label{eq:atomic_spin_transformation}
\end{split}
\end{equation}

Secondly, we introduce the Holstein-Primakoff transformations for effective spin-half systems \cite{Holstein1940}:

\begin{equation}
\begin{split}
    \hat{S}_{+,k} &= \hat{h}_k \\
    \hat{S}_{-,k} &= \hat{h}^{\dagger}_k \\
    \hat{S}_{z,k} &= \left(\frac{1}{2} - \hat{h}^{\dagger}_k\hat{h}_k\right),
    \label{eq:holstein_primakoff_transformation}
\end{split}
\end{equation}

where $\hat{h}$ are the Holstein-Primakoff operators.

\vspace{5mm}

Finally we introduce a transformation between the Holstein-Primakoff operators and the Kittel mode operator. This transformation involves a Fourier transform followed by a Bogoliubov transformation \cite{White2007}. However, for a spherical sample and by ignoring all modes other than the Kittel mode, these transformations reduce to:

\begin{equation}
    \hat{h}_k = \frac{1}{\sqrt{N}}\hat{m}_0.
    \label{eq:mag_transform}
\end{equation}

where $N$ is the number of rare-earth ions in the spherical crystal and $\Hat{m}_0$ is the Kittel mode annihilation operator.

\vspace{5mm}

Applying transformations, Eq.\eqref{eq:atomic_spin_transformation}, Eq.\eqref{eq:holstein_primakoff_transformation} and Eq.\eqref{eq:mag_transform} to Eq.\eqref{eq:adiab_elim_1} we obtain a Hamiltonian in terms of Kittel mode operators:

\begin{equation}
\begin{split}
    \mathcal{H} &= \hbar\delta_{\mu}\Hat{a}^{\dagger}\Hat{a} + \hbar\delta_{o2}\hat{b}^{\dagger}\hat{b} + \hbar\Delta_M\hat{m}_0^{\dagger}\hat{m}_0 \\
    &+ \Big[\hbar\hat{m}_0\left(G^*_{\mu}\hat{a}^{\dagger} - G^*_{o,\Omega}\hat{b}^{\dagger}\right) + \text{H.c.}\Big],
    \label{eq:hamiltonian_magnon_form}
\end{split}
\end{equation}

where we have introduced the parameters,

\begin{align}
    G_{\mu} &= \frac{1}{\sqrt{N}}\sum_kg_{\mu,k}\label{eq:G_mu}\\
    G_{o,\Omega} &= \frac{1}{\sqrt{N}}\sum_k \frac{\Omega_k^*g_{o,k}}{\Delta_{o,k}}\label{eq:G_o}\\
    \Delta_M &= \frac{1}{N}\sum_k\Delta_{\mu,k}
\end{align}

These parameters have a physical interpretation: $G_{\mu}$ describes the coupling between the microwave cavity and the Kittel mode, $G_{o,\Omega}$\ describes the coupling between the optical cavity and the Kittel mode via the two step process $\ket{g} \rightarrow \ket{2} \rightarrow \ket{1}$, and $\Delta_M$ describes the detuning of the microwave field from the Kittel mode transition.

\vspace{5mm}

Working with $G_\mu, G_{o,\Omega} \ll \Delta_M$ we can adiabatically eliminate the Kittel mode from Eq.\eqref{eq:hamiltonian_magnon_form} to obtain:

\begin{equation}
\begin{split}
    \mathcal{H} &= \hbar\delta_{\mu}\Hat{a}^{\dagger}\Hat{a} + \hbar\delta_{o2}\hat{b}^{\dagger}\hat{b} \\
    &-\frac{\hbar}{\Delta_M}|G_{\mu}|^2\hat{a}^{\dagger}\hat{a} - \frac{\hbar}{\Delta_M}|G_{o,\Omega}|^2\hat{b}^{\dagger}\hat{b} \\
    &+ \frac{\hbar}{\Delta_M}\left(G^*_{\mu}G_{o,\Omega}\hat{a}^{\dagger}\hat{b} + G_{\mu}G^*_{o,\Omega}\hat{b}^{\dagger}\hat{a}\right).
    \label{eq:adiab_elim_2}
\end{split}
\end{equation}

Similar to what we saw in Eq.~\eqref{eq:adiab_elim_1}, the third and fourth terms of Eq.~\eqref{eq:adiab_elim_2} represent a shift in the resonance frequency of the two cavities due to the presence of the atoms. These terms can be compensated for in experiment by tuning the cavities (setting $\delta_{\mu} = \frac{\hbar}{\Delta_M}|G_{\mu}|^2$ and $\delta_{o2} =  \frac{\hbar}{\Delta_M}|G_{o,\Omega}|^2$). The third term represents a linear coupling between the microwave and optical cavity fields, with strength $\xi$:
\begin{equation}
    \xi = \frac{G^*_{\mu}G_{o,\Omega}}{\Delta_M}
    \label{eq:micro_to_optic_coupling}
\end{equation}

We have now arrived at a  simplified Hamiltonian for the device, which will be used to estimate the device performance

\begin{equation}
    \mathcal{H} =\hbar\left(\xi\hat{a}^{\dagger}\hat{b} + \xi^*\hat{b}^{\dagger}\hat{a}\right).
    \label{eq:H_simplified}
\end{equation}

\bibliographystyle{apsrev4-1}
\bibliography{bibliography}

\end{document}